\documentclass[12pt,eqsecnum,aps,preprintnumbers,groupedaddress,prd,nofootinbib]{revtex4}

  \usepackage[pdftex]{graphicx}
  \graphicspath{{../pdf/}{../jpeg/}}
 \DeclareGraphicsExtensions{.pdf,.jpeg,.png}
 
\usepackage{amsmath}
\usepackage{enumerate}
\usepackage{amsfonts}
\usepackage{epsfig}

\newcommand{\be}{\begin{equation}}
\newcommand{\ee}{\end{equation}}
\newcommand{\bea}{\begin{eqnarray}}
\newcommand{\eea}{\end{eqnarray}}

\def\lad{L}
\def\le{\left}
\def\ri{\right}
\def\para{g}

\begin{document}

\title {Holographic Lattices, Dimers, and Glasses}

\preprint{NSF-KITP-09-173, SU-ITP-09/42}

\author{Shamit Kachru\footnote{On leave from Department of Physics and SLAC, Stanford University.}}
\affiliation{Kavli Institute for Theoretical Physics and Department of Physics,
University of California,
Santa Barbara, CA 93106, USA\\
{\tt skachru@kitp.ucsb.edu}}

\author{Andreas Karch}
\affiliation{Department of Physics, University of Washington, Seattle, WA 98195-1560, USA\\
{\tt karch@phys.washington.edu}}

\author{Sho Yaida}
\affiliation{Department of Physics, Stanford University, Stanford, CA 94305, USA\\
{\tt yaida@stanford.edu}}


\begin{abstract}
We holographically engineer a periodic lattice of localized fermionic impurities within a plasma medium by putting an array of probe D5-branes in the background produced by $N$ D3-branes.
Thermodynamic quantities are computed in the large $N$ limit via the holographic dictionary.
We then dope the lattice by replacing some of the D5-branes by anti-D5-branes.
In the large $N$ limit, we determine the critical temperature below which the system dimerizes with bond ordering.
Finally, we argue that for the special case of a square lattice our system is glassy at large but finite $N$, with the low temperature physics dominated by a huge collection of metastable dimerized configurations without long-range order, connected only through tunneling events.
\end{abstract}

\maketitle

\section{Introduction}\label{introduction}

Recently there has been a flurry of activity applying the anti-de Sitter (AdS)/conformal field theory (CFT) correspondence~\cite{MaldacenaOriginal,GKP,Witten} to the study of condensed matter systems.   Holographic systems where the CFT exhibits strongly coupled avatars of metallic phases \cite{Karch}, superfluid-insulator transitions \cite{SI}, superconductivity~\cite{Gubser,HHH}, and Fermi liquid theory \cite{Parnachev,Sungsik,John,Zaanen} have all been proposed.
Also proposed are gravity duals for scale-invariant nonrelativistic field theories, enjoying Galilean invariance \cite{Son,BM} or the smaller Lifshitz symmetry group \cite{Shamit}.
For reviews, see~\cite{Hartnoll,Herzog,McGreevy}.
One of the driving forces behind such vigorous activity is possible applications to condensed matter systems with quantum critical points, such as heavy fermion materials and possibly cuprate superconductors.
The AdS/CFT correspondence itself may even be thought of as a prime example of quantum critical phenomena where, at the conformal fixed point of a family of field theories, there appear emergent gravitons.
At a more concrete level, one significant advantage the AdS/CFT correspondence offers over other methods of analyzing model field theories is the ability to compute real-time correlators, and hence to gain a handle on transport properties, in strongly coupled field theories.
For a review of this aspect with relevant references, see \cite{Sonreview}.

One limitation of the studies to date is that the systems considered so far in the literature are spatially homogeneous.  While such homogeneous systems might suffice for studies of critical phenomena, where the correlation length diverges and the microscopic structure of solids does not play any role, these holographic toy models are completely inadequate when it comes to questions involving the underlying lattice structure of condensed matter systems. In the first part of this paper, we attempt to remedy this situation by explicitly constructing holographic systems endowed with periodic lattice structures, in the context of type IIB string theory.

To this end, we first consider a probe D5-brane in the  ${\rm AdS}_5\times S^5$ background geometry, wrapping ${\rm AdS}_2\times S^4$. From the boundary field theory's point of view, this corresponds to adding localized fermionic degrees of freedom residing on a pointlike impurity coupled to ${\cal N}=4$ supersymmetric $SU(N)$ gauge theory.
This is a particular example of the more general structure of defect conformal field theory,
investigated in the AdS/CFT context in,
for example,~\cite{Karch:2000gx,defectone,defecttwo}.
Placing an array of such D5-branes in the asymptotically ${\rm AdS}_5\times S^5$ black brane geometry, we obtain a lattice of impurities immersed in the ${\cal N}=4$ plasma medium at finite temperature~\cite{WittenThermal}. Using the holographic dictionary, we can compute thermodynamic quantities of this system in the large $N$ and large 't Hooft coupling limit, where the gravitational description is accurate.

We then proceed further and dope the system, replacing half (say) of the D5-branes by anti-D5-branes. This doping introduces an interesting phenomenon of dimerization: whereas both D5- and anti-D5-branes go straight down into the black brane horizon at high temperature, at low temperature they pair up by connecting with each other far from the horizon. Again, in the large $N$ and large 't Hooft coupling limit, we can compute the free energy and determine the critical temperature at which such dimerization occurs. With ease, we can also engineer systems with various plateaux of temperature ranges within which only a fraction of the unit cell of the lattice dimerizes. In the special case of a square lattice, we argue that our system exhibits glassy behavior at large but finite $N$.

The organization of the paper is as follows.
In Sec.\ref{lattice}, we build a holographic lattice and study its thermodynamic properties.
Section \ref{pair} is an intermezzo where we consider a pair of a D5-brane and an anti-D5-brane in the black brane background.
Equipped with the result in Sec.\ref{pair}, in Sec.\ref{dimer}, we engineer various systems which dimerize at low temperature, and determine the critical dimerization temperature(s).
In Sec.\ref{glass}, we briefly argue that a special case of our construction yields a glassy system.
We conclude by suggesting numerous future directions in Sec.\ref{conclusion}. Appendix A summarizes our conventions.

\section{Holographic Lattices}\label{lattice}

In this section, we gradually work up to a lattice of probe D5-branes in the black brane background, while explaining its field theory dual and calculating associated thermodynamic quantities.
The gravitational side of our holographic systems is governed by the type IIB supergravity action plus a probe D5 action for each D5-brane, $S=S_{\rm IIB}+\sum_{i}S_{\rm D5_{\it i}}$ (see Appendix A for details). For simplicity, we will not consider coincident D5-branes in this paper, and forces between D5-branes, which vanish at zero temperature as the configuration is Bogomol'nyi-Prasad-Sommerfield (BPS), but would be induced
at finite temperature, will not matter at the order to which we work.

\subsection{Black brane background}

The type IIB supergravity is well known to have the following asymptotically ${\rm AdS}_5\times S^5$ black brane solution:
\bea
B_{(2)}&=&C_{(0)}=C_{(2)}=0,\ \ \frac{e^{\Phi}}{g_s}=1,\\
g_{\mu\nu}dx^{\mu}dx^{\nu}&=&-f(r)dt^2+\frac{dr^2}{f(r)}+\frac{r^2}{\lad^2}(\sum_{i=1}^3 dx_i^2)+\lad^2 (d\theta^2+{\rm sin}^2\theta d\Omega_4^2)\\
&&{\rm with}\ \  f(r)=\frac{r^2}{\lad^2}\left(1-\frac{r_{+}^4}{r^4}\right)\ \ {\rm and}\ \ \lad^4\equiv4\pi g_sN \alpha'^2,\\
F_{(5)}&=&dC_{(4)}=-(1+*)\frac{4r^3}{\lad^4}(dt\wedge dr \wedge dx_1 \wedge dx_2 \wedge dx_3).
\eea
For later convenience, we foliated $S^5$ by $S^4$s parametrized by their latitudes $\theta$, and $d\Omega_4^2$ is the metric on a unit 4-sphere.
This solution can be obtained as the decoupling limit of the geometry generated by a stack of
$N$ nonextremal D3-branes and, as such, it describes ${\cal N}=4$ supersymmetric $SU(N)$ gauge theory at finite temperature, with the Yang-Mills (YM) coupling constant of the gauge theory given in terms of the string coupling constant as $g_{\rm YM}^2=4\pi g_s$~\cite{WittenThermal}.

The field theory's temperature is identified with the Hawking temperature of the geometry above, namely, $T(r_{+})=\frac{1}{2\pi}[\frac{1}{\sqrt{g_{rr}}}\frac{d}{dr}\sqrt{-g_{tt}}]|_{r=r_{+}}
=\frac{r_{+}}{\pi\lad^2}$. The free energy $F[T]$ of the macroscopic configuration can be computed through the holographic dictionary, which reads in the classical limit as
\bea
e^{-\frac{1}{T}F[T]}=Z[T]=e^{-I[T]}.
\eea
Here, $I[T]$ is the properly dimensionally reduced $(4+1)$-dimensional Euclidean action, evaluated for the configuration with temperature $T$.
This is by now a standard practice and the result is
\be
F_{\rm IIB}[T(r_{+})]=-\frac{1}{128}V_3\le(\frac{r_{+}^4}{\pi^4\lad^8}\ri) \le(\frac{\lad^8}{g_s^2\alpha'^4}\ri)=-\frac{\pi^2}{8}V_3T^4N^2,
\ee
where $V_3$ denotes the 3-dimensional volume in ${\overrightarrow x}$ direction.

\subsection{D5 embedding}\label{single}
Let us now place a single D5-brane in the black brane background above. There is a one-parameter family of analytic solutions for static D5 embeddings, going straight down into the black brane, characterized by latitudes $\theta_{\nu}$ of $S^4$ that they wrap. The position ${\overrightarrow a}$ in the ${\overrightarrow x}$ directions and the direction of the $S^4$ within $S^5$ give other parameters as well. But for a single D5-brane, we can take these to be fixed at any value without loss of generality, whereas for multiple D5-branes to be considered below, their relative values would matter.  Explicitly, by parametrizing the 4-spheres by $(\theta_1, \theta_2, \theta_3, \phi_4)$ and denoting the  coordinates on the D5-brane by $\xi=(\tau, \rho, \theta_1, \theta_2, \theta_3, \phi_4)$, the embeddings are given by\footnote{We choose the orientation of the spacetime so that ${\epsilon}_{tr123\theta\theta_1\theta_2\theta_3\phi_4}=+\sqrt{-g}$ and those of D5-branes so that ${\epsilon}_{\tau\rho\theta_1\theta_2\theta_3\phi_4}=+\sqrt{-g^{({\rm induced})}}$.}
\be
(\tau, \rho, \theta_1, \theta_2, \theta_3, \phi_4)\mapsto(\tau,  \rho, {\overrightarrow 0}, \theta_{\nu}, \theta_1, \theta_2, \theta_3, \phi_4)\ \ {\rm with}\ \  \rho\in(0,\infty)\ \ {\rm and}\ \ \tau\in(-\infty, \infty),
\ee
where the worldvolume flux depends on $\theta_\nu$, and is given by
\be
(2\pi\alpha'{\cal F})_{\tau\rho}=-(2\pi\alpha'{\cal F})_{\rho\tau}={\rm cos}\theta_{\nu}\ \ {\rm with\ all\ other\ components\ vanishing}.\ \
\ee
Actually, as explained in~\cite{Stable}, possible values of the latitude $\theta_{\nu}\in(0,\pi)$ are quantized: defining a parameter $\nu\in(0,1)$ by
\be\label{rel1}
\nu\equiv\frac{1}{\pi}\le(\theta_{\nu}-{\rm sin}\theta_{\nu}{\rm cos}\theta_{\nu}\ri),
\ee
we have the quantization condition
\be\label{rel2}
n\equiv\nu N\in {\mathbb Z}.
\ee

In the dual field theory, introducing such a D5-brane in the bulk corresponds to coupling localized massless fermions to the ${\cal N}=4$ gauge theory as
\be
S_{\rm field\ theory}=S_{{\cal N}=4}+\int dt \le[i\chi^{\dagger}_b\partial_t\chi^b+\chi^{\dagger}_b\le\{(A_0(t, {\overrightarrow 0}))^b_c+v^I(\phi_I(t, {\overrightarrow 0}))^b_c\ri\}\chi^c\ri],
\ee
where $S_{{\cal N}=4}$ is the action for ${\cal N}=4$ supersymmetric $SU(N)$ gauge theory~\cite{WeinbergIII}, $A_{\mu}$ and $\phi_{I}$ are its gauge and scalar fields in the adjoint representation of $SU(N)$, respectively, and $v^{I}$ is a unit 6-vector determined by the direction of the $S^4$ within $S^5$ wrapped by the D5-brane.
The parameter $n\equiv\nu N$, on the other hand, determines the number of fermions at the site ${\overrightarrow 0}$~\cite{Ensemble}. More precisely, it corresponds to taking the ensemble with the density matrix
\be
\frac{n!(N-n)!}{N!}\sum_{b_1<...<b_n}\mid b_1,...,b_n\rangle  \langle b_1,...,b_n\mid
\ee
where, schematically,\footnote{The states as written in~Eq.(\ref{schematic}) are not invariant under small gauge transformations, and they need to be supplemented by nonlocal operators for any nonzero coupling.}
\be\label{schematic}
\mid b_1,...,b_n\rangle=\chi^{\dagger}_{b_1}...\chi^{\dagger}_{b_n}\vert 0\rangle.
\ee

Now we would like to compute the D5-brane contribution to the free energy.
Turning it into the appropriate Euclidean configuration, the Euclidean action of the D5-brane is evaluated to be
\cite{Hartnoll:2006is,
Gomis:2006sb,Yamaguchi:2006tq}\footnote{To get correct results for free energies, it is very crucial to add the appropriate surface term for the D5 action, analogous to the Gibbons-Hawking surface term for the Einstein-Hilbert action, so that we can consistently impose a Dirichlet boundary condition at $r=r_{\rm cutoff}$.}
\bea
I_{\rm D5}^{({\rm bulk})}&=&\frac{\lad^4  {\rm sin}^3\theta_{\nu}{\rm vol}(S^4)}{(2\pi)^5g_s \alpha'^3}\frac{1}{T}(r_{\rm cutoff}-r_{+}),
\eea
where we regulated it with the regulator $r_{\rm cutoff}$. We need to holographically renormalize this quantity.
To this end, we subtract off the Euclidean action of the analogous D5 embedding into pure ${\rm AdS}_5\times S^5$ spacetime,
\be
I_{\rm D5}^{({\rm counter})}=\frac{\lad^4 {\rm sin}^3\theta_{\nu}{\rm vol}(S^4)}{(2\pi)^5 g_s \alpha'^3}\frac{1}{T'(T)}(r_{\rm cutoff}),
\ee
where we take its temperature to be $T'(T)$ given by
\be
\frac{1}{T'(T)}\sqrt{\left(\frac{r_{\rm cutoff}^2}{\lad^2}\right)}=\frac{1}{T}\sqrt{\left(\frac{r_{\rm cutoff}^2}{\lad^2}\right)\left(1-\frac{r_{+}^4}{r_{\rm cutoff}^4}\right)}
\ee
so that the induced metric on the surface at $r_{\rm cutoff}$ is the same as in the relevant black brane configuration~\cite{WittenThermal}. This procedure is justified in the context of holographic renormalization~\cite{Skenderis} because both configurations are the solutions to the same equations of motion with the same leading Dirichlet boundary condition at $r=r_{\rm cutoff}$.
Thus, the renormalized D5 Euclidean action is given by
\be
I_{\rm D5}^{({\rm renormalized})}=\lim_{r_{\rm cutoff}\rightarrow\infty}(I_{\rm D5}^{({\rm bulk})}-I_{\rm D5}^{({\rm counter})})=-\frac{\lad^4   {\rm sin}^3\theta_{\nu}{\rm vol}(S^4)}{(2\pi)^5g_s \alpha'^3}\frac{r_{+}}{T}.
\ee
Hence the leading D5 contribution to the free energy is
\begin{equation}
F_{\rm D5}=T I_{\rm D5}^{({\rm renormalized})}=-\frac{\lad^4{\rm sin}^3\theta_{\nu}{\rm vol}(S^4)}{(2\pi)^5g_s \alpha'^3}r_{+}=-\frac{ {\rm sin}^3\theta_{\nu}}{3\pi } N \sqrt{\lambda} T,
\end{equation}
where we introduced the 't Hooft coupling $\lambda\equiv g_{\rm YM}^2 N$.

\subsection{Lattice of D5-branes}
Finally, let us consider an array of D5-branes. For simplicity, we consider a cubic lattice with
lattice spacing $a$, and let the D5-branes wrap $S^4$s of the same direction in $S^5$ and
at the same latitude.\footnote{This guarantees the supersymmetry of the zero temperature configuration.}
On the boundary field theory side, we are just constructing a cubic lattice of impurities.
Assembling the results of preceding subsections, the free energy of the system is given by
\be\label{freelattice}
F[T]=V_3\le[-N^2\le(\frac{\pi^2}{8}T^4+...\ri)-\sqrt{\lambda}N\le(\frac{ {\rm sin}^3\theta_{\nu}}{3\pi }\frac{1}{a^3}T+...\ri)\ri],
\ee
and hence the entropy is
\be\label{entropylattice}
S[T]=-\frac{dF}{dT}=V_3\le[N^2\le(\frac{\pi^2}{2}T^3+...\ri)+\sqrt{\lambda}N\le(\frac{ {\rm sin}^3\theta_{\nu}}{3\pi }\frac{1}{a^3}+...\ri)\ri],
\ee
where dots in each parenthesis indicate $\frac{1}{\sqrt{\lambda}}$ suppressed terms coming from stringy corrections in the bulk language. When expressed in terms of boundary field theory's quantities specifying bulk boundary conditions, as above, Eqs.(\ref{freelattice}) and (\ref{entropylattice}) capture the whole leading contributions in the large $N$ and large 't Hooft limit, up to order $N^1$. In particular, backreaction of D5-branes onto the geometry mingles only into the order $N^0$ contribution together with other bulk quantum effects.
For a formal argument within the context of Einstein gravity and toy examples of how this works, see~\cite{Proof}.\footnote{The basic idea is that one expands the full solution to the coupled supergravity and probe action in a power series in $\frac{1}{N}$ and then substitutes this expansion back into the action to obtain the free energy.
At order $N^1$ there are two potential contributions: the probe action evaluated on the leading solution and the correction to the contribution from the supergravity action due to the $\frac{1}{N}$ correction in the background geometry.
The latter however vanishes due to the equations of motion obeyed by the background.}

The formulas above obtained through the bulk geometric analysis, however, have a
limited range of validity:
at extremely low temperature (small horizon radius), our geometric analysis breaks down.
This happens when the entropy contribution of the D5-branes is comparable to that of the background black brane geometry, namely, when $T^3a^3\sim\frac{\sqrt{\lambda}}{N}$. The reason behind
this breakdown is that the backreaction of D5-branes at the horizon of the black brane cannot be treated as small perturbation any more, since the density of D5-branes at the horizon, $\frac{\lad^3}{r_{+}^3 a^3}$, increases as $r_{+}$ decreases.
Hence our formulas above are not applicable outside the regime
\be
T a>>\le(\frac{\sqrt{\lambda}}{N}\ri)^{\frac{1}{3}}.
\ee

Nevertheless, there is one ``thermodynamic" quantity which we can evaluate at zero temperature, namely, the entropy.
This comes from the fact that our lattice preserves supersymmetry at zero temperature, and getting the entropy is tantamount to counting the number of BPS configurations with given charges.
Carrying out the analysis, the entropy at zero temperature is
\be
S[T=0]={\rm ln}\le\{\le(\frac{N!}{n!(N-n)!}\ri)^{\frac{V_3}{a^3}}\ri\}=\le(\frac{V_3}{a^3}\ri){\rm ln}\le\{\le(\frac{N!}{n!(N-n)!}\ri)\ri\},
\ee
which is applicable for any $N$ and $g_{\rm YM}$, and it asymptotes to $\le(\frac{V_3}{a^3}\ri)\times N\times \{-\nu{\rm ln}\nu-(1-\nu){\rm ln}(1-\nu)\}$ for large $N$.\footnote{The large entropy at zero temperature is often thought to be ``unphysical." Here we interpret it just as a peculiarity of highly supersymmetric systems.}
This should be contrasted with the finite temperature results which
yields the leading D5 contribution to the entropy of order $N\sqrt{\lambda}$, enhanced by $\sqrt{\lambda}$. This is analogous to the strong coupling enhancement of fundamental matter in the deconfined plasma observed in~\cite{D7}, as opposed to the zero temperature conformal medium.

\section{Intermezzo: D5/anti-D5 pair}\label{pair}
In this intermezzo, we consider a pair of a D5-brane and an anti-D5-brane immersed in the black brane background. In the bulk gravitational description, the anti-D5-brane differs from the D5-brane in its orientation, which manifests in a reversal of sign of its coupling to the background Ramond-Ramond fields. On the boundary field theory side, localized fermions introduced by the anti-D5-brane transform in the antifundamental representation of $SU(N)$ while those introduced by the D5-brane transform in the fundamental representation.\footnote{We work in a convention in which the relation between the number of fermions introduced by the anti-D5-brane ${\bar n}={\bar \nu}N$ and its latitude ${\bar \theta}_{\bar \nu}$ is given by Eqs.(\ref{rel1}) and (\ref{rel2}).}

We will consider one D5-brane placed at ${\overrightarrow x}=(+\frac{\Delta x}{2},0,0)$ and one anti-D5-brane placed at $(-\frac{\Delta x}{2},0,0)$, both of them wrapping $S^4$s of the same angle and latitude as $r\rightarrow\infty$. Intuitively (see Fig.1), the ``ends" of both the D5- and anti-D5-branes are sucked into the black brane when the horizon radius $r_{+}$ is large (or the separation $\Delta x$ is large), whereas they reconnect outside the horizon for small horizon radius (or small separation).

\begin{figure}[t]
\includegraphics[scale=0.65,angle=0]{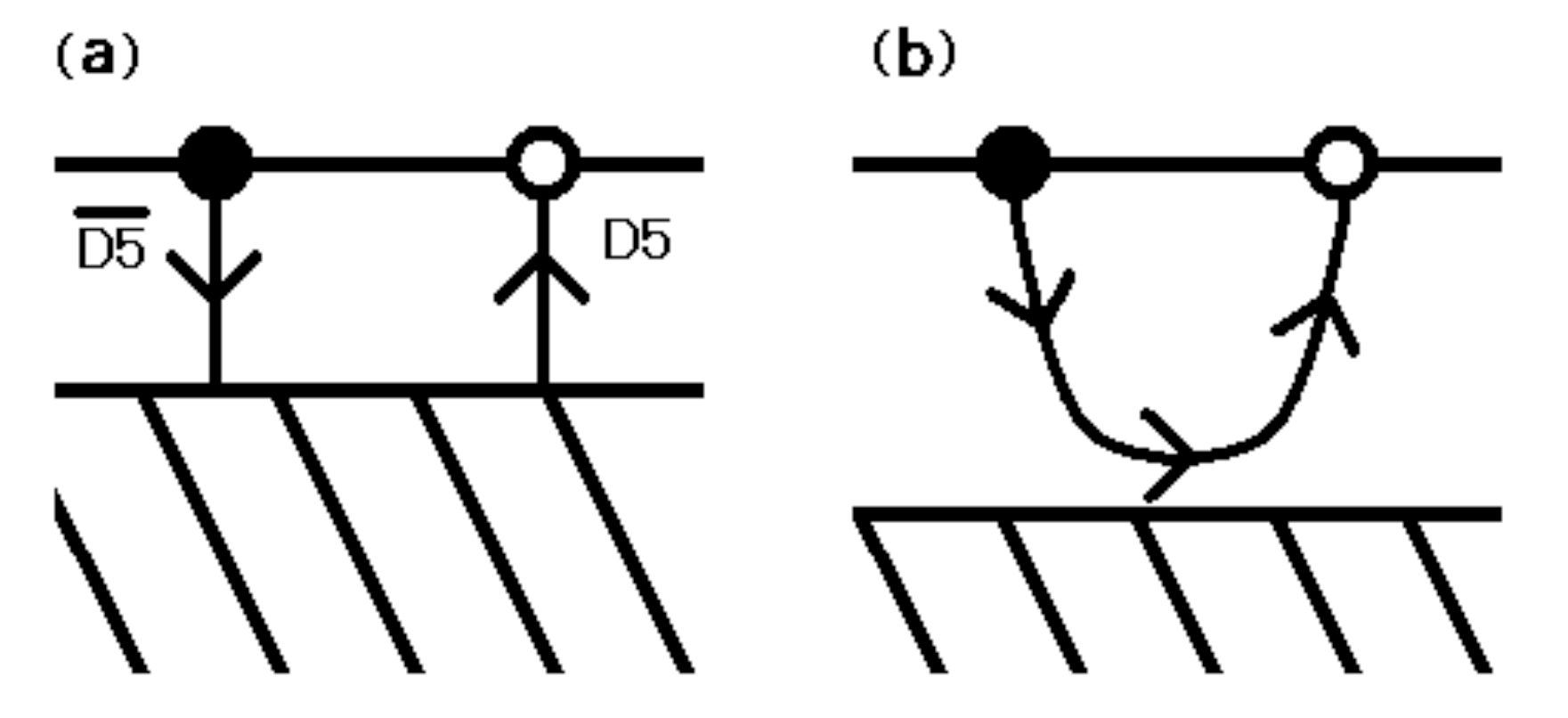}
\caption{Predimerization transition. (a)Disconnected configuration dominates at high temperature. (b)Connected configuration dominates at low temperature.}
 \label{1reconnection}
\end{figure}

\subsection{Disconnected configuration}
The obvious candidate stable configuration of such a pair is just two separated configurations of the sort considered in Sec.\ref{single} with $\theta_{\nu}={\bar \theta}_{\bar \nu}$ [see Fig.1(a)]. Its free energy is just twice that of the single D5-brane:
\be
F_{\rm D5}+F_{{\bar {\rm D5}}}=-\le(2\frac{\lad^4 {\rm sin}^3\theta_{\nu}{\rm vol}(S^4)}{(2\pi)^5g_s\alpha'^3}r_{+}\ri) .
\ee
Note that it is independent of the separation $\Delta x$.

\subsection{Connected configuration}
Another candidate solution with the given boundary condition is a reconnecting solution [see Fig.1(b)]: a reconnecting D5-brane starts at $r=\infty$ with ${\overrightarrow x}=(-\frac{\Delta x}{2},0,0)$, dips into the bulk, and then comes back to $r=\infty$ now with ${\overrightarrow x}=(+\frac{\Delta x}{2},0,0)$, effectively reversing its orientation as it should.\footnote{Incidentally, this is the reason why a pair of two D5-branes cannot reconnect.} Explicitly, we have
\be
\theta(\rho)=\theta_{\nu} \ \ {\rm and}\ \ (2\pi\alpha'{\cal F})_{\tau\rho}={\rm cos}\theta_{\nu}\frac{1}{\sqrt{1-\frac{1}{{\rm sin}^6\theta_{\nu}}\le(\frac{\lad^4 k^2}{r^4(\rho)-r_{+}^4}\ri)}}\sqrt{\le(\frac{\partial r}{\partial\rho}\ri)^2},
\ee
and now ${\overrightarrow x}$ is also a function of the coordinate $\rho$ on the D5-brane.
As $r(\rho)$ starts from $\infty$ and goes toward the turning point $r_{\rm turn}=\{ r_{+}^4+\frac{\lad^4 k^2}{{\rm sin}^6\theta_{\nu}}\}^{\frac{1}{4}}$,
\be
{\overrightarrow x}(\rho)=-(k,0,0)\times\int_{r_{\rm turn}}^{r(\rho)} dr'\frac{\lad^4}{r'^4-r_{+}^4}\frac{1}{\sqrt{{\rm sin}^6\theta_{\nu}-\frac{\lad^4 k^2}{r'^4-r_{+}^4}}},
\ee
while as $r(\rho)$ goes back up from $r_{\rm turn}$ and ends at $\infty$,
\be
{\overrightarrow x}(\rho)=+(k,0,0)\times\int_{r_{\rm turn}}^{r(\rho)} dr'\frac{\lad^4}{r'^4-r_{+}^4}\frac{1}{\sqrt{{\rm sin}^6\theta_{\nu}-\frac{\lad^4 k^2}{r'^4-r_{+}^4}}}.
\ee
Here the parameter $k$ is implicitly determined through
\be
\frac{r_{+}}{\lad^2}\Delta x=\left[2{\tilde k} \int_{(1+{\tilde k}^2)^{\frac{1}{4}}}^{\infty}dz\sqrt{\frac{1}{(z^4-1)(z^4-1-{\tilde k}^2)}}\right]\ \ {\rm with}\ \ {\tilde k}\equiv\frac{\lad^2}{r_{+}^2{\rm sin}^3\theta_{\nu}}k.
\ee
The function on the right-hand side is plotted as a function of ${\tilde k}$ in Fig.2. Note that, for a fixed distance, such a solution does not exist for large enough $r_{+}$. Also, for small enough $r_{+}$, there are two distinct solutions with different ${\tilde k}$. By computing their free energies, we will determine which configuration among the two is dominant.

\begin{figure}[t]
\includegraphics[scale=0.65,angle=0]{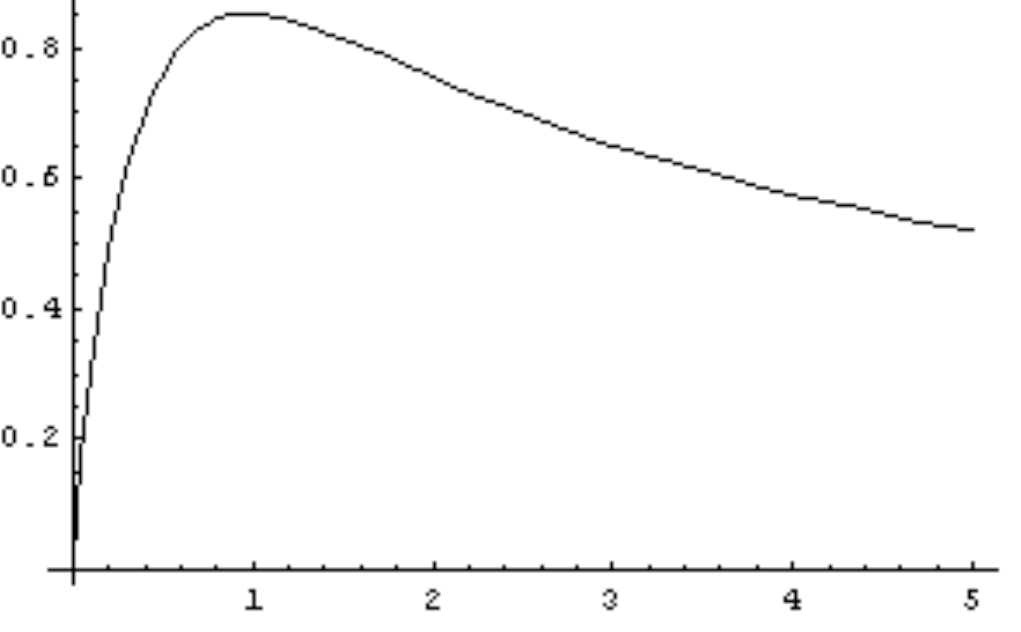}
\caption{$\frac{r_{+}}{\lad^2}\Delta x$ as a function of ${\tilde k}$.}
 \label{2distance}
\end{figure}

\begin{figure}[t]
\includegraphics[scale=0.65,angle=0]{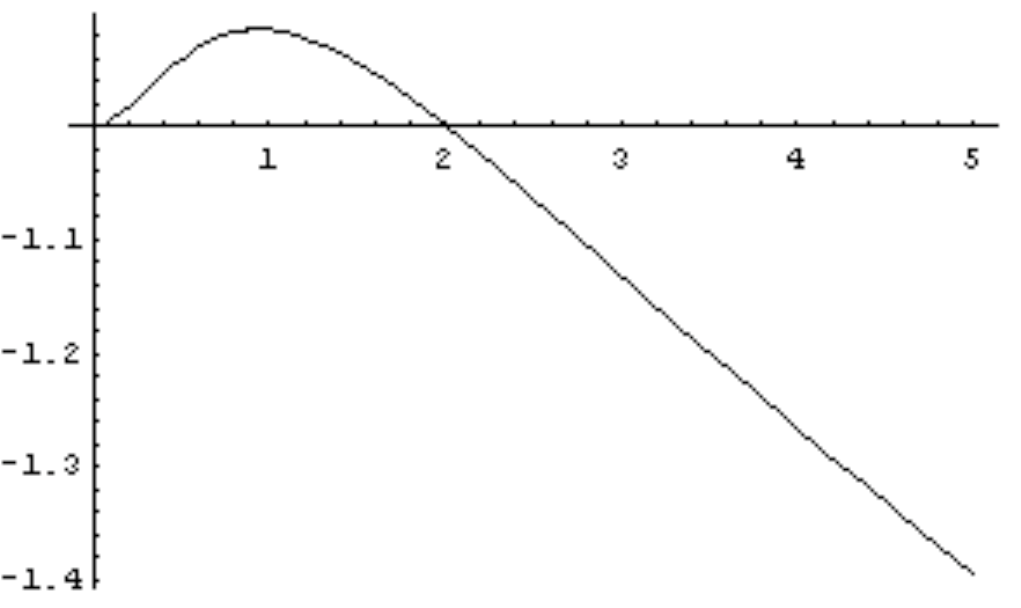}
\caption{${\tilde F}_{\rm connected}({\tilde k})$ as a function of ${\tilde k}$.}
 \label{3free}
\end{figure}

We can follow the same regularization procedure as presented in Sec.\ref{single} to get the free energy of the reconnecting solution, which yields
\bea
F_{\rm connected}&=&\left(2\frac{\lad^4 {\rm sin}^3\theta_{\nu}{\rm vol}(S^4)}{(2\pi)^5g_s \alpha'^3}r_{+}\right)\left[-(1+{\tilde k}^2)^{\frac{1}{4}}+\int_{(1+{\tilde k}^2)^{\frac{1}{4}}}^{\infty}dz\left\{\sqrt{\frac{z^4-1}{z^4-1-{\tilde k}^2}}-1\right\}\right]\ \ \ \\
&\equiv&\left(2\frac{\lad^4 { {\rm sin}^3\theta_{\nu}\rm vol}(S^4)}{(2\pi)^5g_s \alpha'^3}r_{+}\right) {\tilde F}_{\rm connected}({\tilde k}).\nonumber
\eea
The function ${\tilde F}_{\rm connected}({\tilde k})$ is plotted in Fig.3.
Comparing with Fig.2, we see that when $r_{+}$ is low enough to allow two solutions for a given $\Delta x$, the configuration with higher ${\tilde k}$, and hence the one that goes less deeply into the bulk, is the dominant configuration among the two since it has smaller free energy. All the calculations in this section are virtually identical to the discussion of Wilson lines at finite temperature from fundamental strings in AdS spacetime~\cite{Rey:1998bq,Brandhuber:1998bs}, the only difference being the overall prefactor which for us is set
by the effective string tension of the wrapped D5-brane, whereas for the Wilson lines it was set by the fundamental string tension.

\subsection{Predimerization transition}
Here comes the punch line of this intermezzo: for a fixed separation $\Delta x$, there is a large $N$ phase transition as we decrease $r_{+}$.
To see this, let us start in the regime where $\frac{r_{+}\Delta x}{\lad^2}$ is very large. In such a regime, as we noted earlier, there is no reconnecting solution and thus the disconnected configuration dominates [Fig.4(a)]. As we lower $r_{+}$, there appear two reconnecting solutions, but at the onset they both have free energies higher than that of the disconnected solution and thus stay subdominant [Fig.4(b)]. As we lower $r_{+}$ further and reach the value ${\tilde k}={\tilde k}_c\approx 2$ or
\be
\frac{r_{+}\Delta x}{\lad^2}\approx 0.7,
\ee
we see one of the reconnecting solutions, namely, the one which goes less deeply into the bulk, emerges as the most dominant configuration [Fig.4(c)]. Noting that free energies of the D5-branes have a multiplicative factor $N$, we see that this marks a phase transition in the $N=\infty$ limit.

\begin{figure}[t]
\includegraphics[scale=0.54,angle=0]{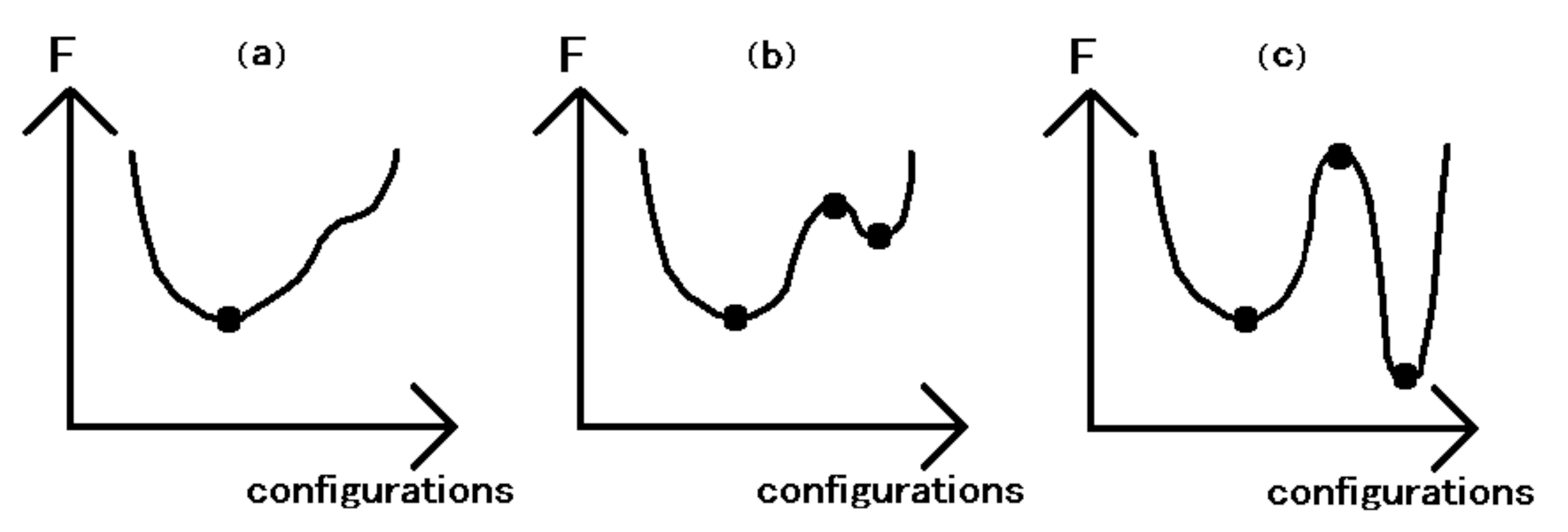}
\caption{Free energies as functionals of D5 configurations at various temperature ranges.
(a)At high temperature, there is only one extremum corresponding to the disconnected solution.
(b)At intermediate temperature, there appear two new local extrema corresponding to two connected solutions.
(c)At low temperature, the stable connected solution dominates.}
 \label{4transition}
\end{figure}

This $N=\infty$ phase transition will be rendered to be a crossover for large but finite $N$.
We may avoid this by building a (3+1)-dimensional lattice whose unit cell has pairs of D5- and anti-D5-branes. We expect this system to have a genuine first-order phase transition in the thermodynamic limit $V_3\rightarrow\infty$, surviving a journey away from the $N=\infty$ limit.

\section{Holographic Dimers}\label{dimer}
Equipped with the result in the previous section, we can now build holographic dimers, systems which dimerize at low temperature. Dimer systems are extensively studied in the condensed matter physics community, especially as a part of attempts in explaining cuprate superconductivity, and
in trying to get a handle on the physics of Mott insulators \cite{Sachdev}. And since the physics responsible for cuprate superconductivity is believed to be primarily (2+1)-dimensional (justifying the use of a Lawrence-Doniach model)~\cite{Leggett}, such studies have focused on (weakly interacting layers of) (2+1)-dimensional dimer systems. To keep the close analogy with such studies, we will focus on (2+1)-dimensional lattices, though it should be kept in mind that we are thinking of eventually stacking such (2+1)-dimensional lattices so that we can evade the Coleman-Mermin-Wagner-Hohenberg theorem and have a phase transition at finite temperature.

\subsection{Dimerization through bond ordering}\label{VBS}

\begin{figure}[t]
\includegraphics[scale=0.65,angle=0]{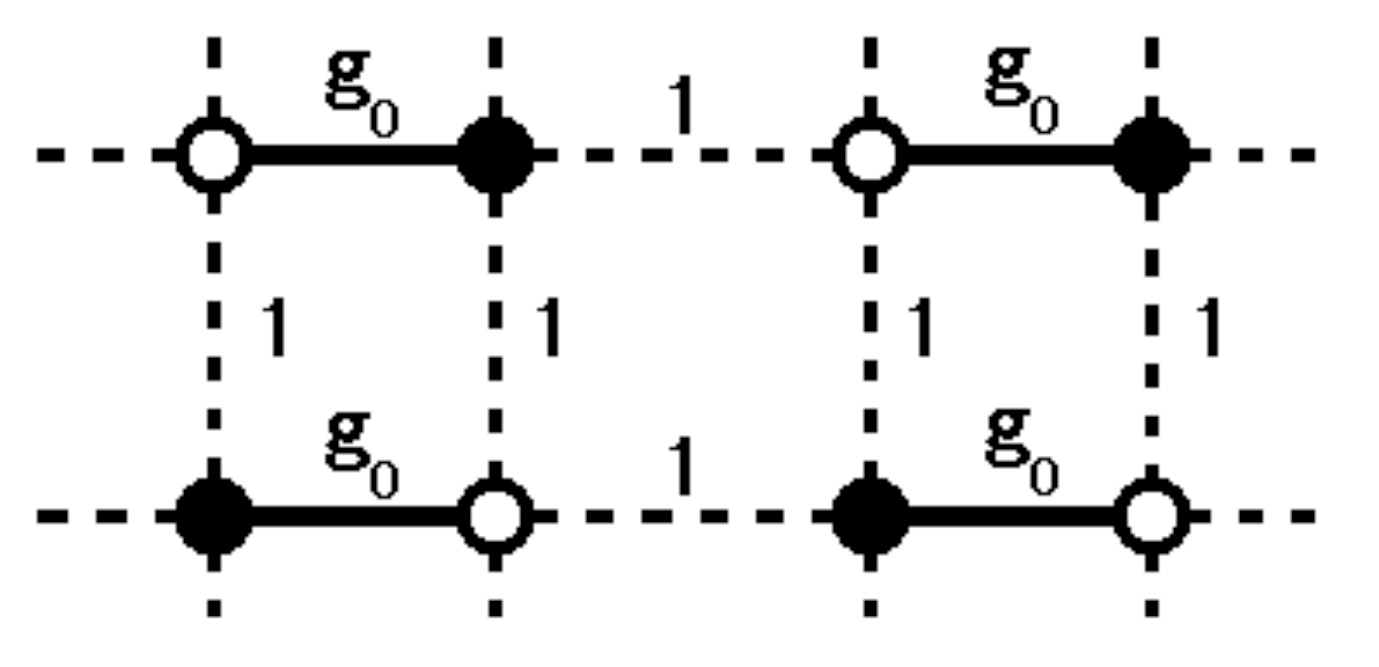}
\caption{A simple model that experiences the phase transition analogous to the N\'eel-VBS phase transition.}
 \label{5VBS}
\end{figure}

Let us consider the lattice (see Fig.5) in which D5-branes are placed at
\be
a\le\{\le(-\frac{\para_0}{2}, +\frac{1}{2}\ri)+\le((1+\para_0)n_x, 2n_y\ri)\ri\}\ \ {\rm and}\ \ a\le\{\le(+\frac{\para_0}{2}, -\frac{1}{2}\ri)+\le((1+\para_0)n_x, 2n_y\ri)\ri\}
\ee
and anti-D5-branes are placed at
\be
a\le\{\le(+\frac{\para_0}{2}, +\frac{1}{2}\ri)+\le((1+\para_0)n_x, 2n_y\ri)\ri\}\ \ {\rm and}\ \ a\le\{\le(-\frac{\para_0}{2}, -\frac{1}{2}\ri)+\le((1+\para_0)n_x, 2n_y\ri)\ri\}
\ee
with $(n_x, n_y)\in{\mathbb Z}^2$ and $0<\para_0<1$.
Working in the large $N$ and large 't Hooft coupling limit, then, as we cool down to $\frac{r_{+}a\para_0}{\lad^2}\approx 0.7$ or
\be
T_c\approx \frac{0.2}{a\para_0},
\ee
our lattice dimerizes through bubble nucleation of dimerized regions. This transition is analogous to the N\'eel-VBS (valence bond solid) phase transition~\cite{Sachdev}, though here it is a first-order phase transition as a function of temperature, not a quantum critical second-order phase transition at zero temperature. Keeping $\frac{\sqrt{\lambda}}{N}<<1$, this transition happens away from the regime where backreaction becomes important, justifying our geometric analysis.

\subsection{Plateaux}

\begin{figure}[t]
\includegraphics[scale=0.65,angle=0]{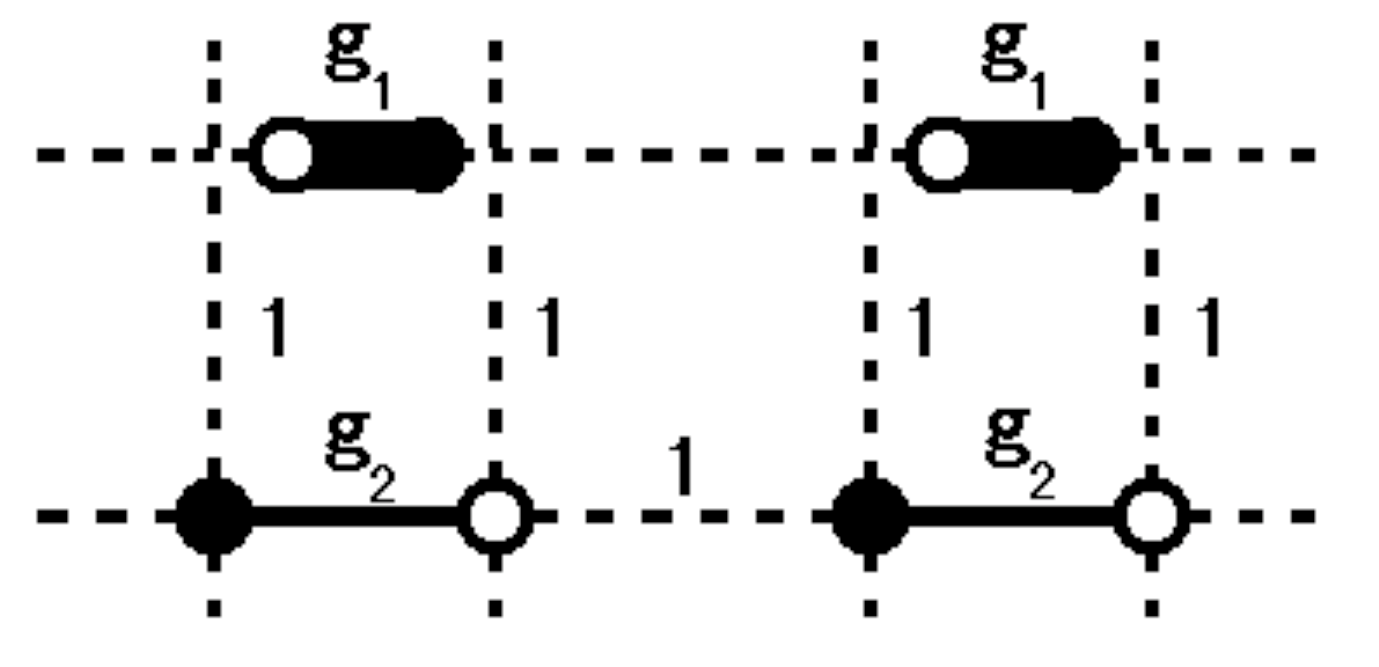}
\caption{A model with a plateau of temperature range within which only half of each unit cell dimerizes.}
 \label{6plateau}
\end{figure}

To illustrate how easy it is to engineer variants of the simple VBS dimerization described above,
let us now consider the lattice (see Fig.6) in which D5-branes are placed at
\be
a\le\{\le(-\frac{\para_1}{2}, +\frac{1}{2}\ri)+\le((1+\para_2)n_x, 2n_y\ri)\ri\}\ \ {\rm and}\ \ a\le\{\le(+\frac{\para_2}{2}, -\frac{1}{2}\ri)+\le((1+\para_2)n_x, 2n_y\ri)\ri\}
\ee
and anti-D5-branes are placed at
\be
a\le\{\le(+\frac{\para_1}{2}, +\frac{1}{2}\ri)+\le((1+\para_2)n_x, 2n_y\ri)\ri\}\ \ {\rm and}\ \ a\le\{\le(-\frac{\para_2}{2}, -\frac{1}{2}\ri)+\le((1+\para_2)n_x, 2n_y\ri)\ri\}
\ee
with $(n_x, n_y)\in{\mathbb Z}^2$ and $0<\para_1<\para_2<1$.
Then, again working in the large $N$ and large 't Hooft coupling limit, as we cool down to $T_{1c}\approx \frac{0.2}{a\para_1}$, only a half of each unit cell dimerizes, while as we cool further down to $T_{2c}\approx \frac{0.2}{a\para_2}$, the entire cell dimerizes.
In this way, we can engineer a lattice with plateaux of temperature ranges within each of which only a fraction of each unit cell dimerizes.

\section{Holographic Glasses}\label{glass}

\begin{figure}[t]
\includegraphics[scale=0.65,angle=0]{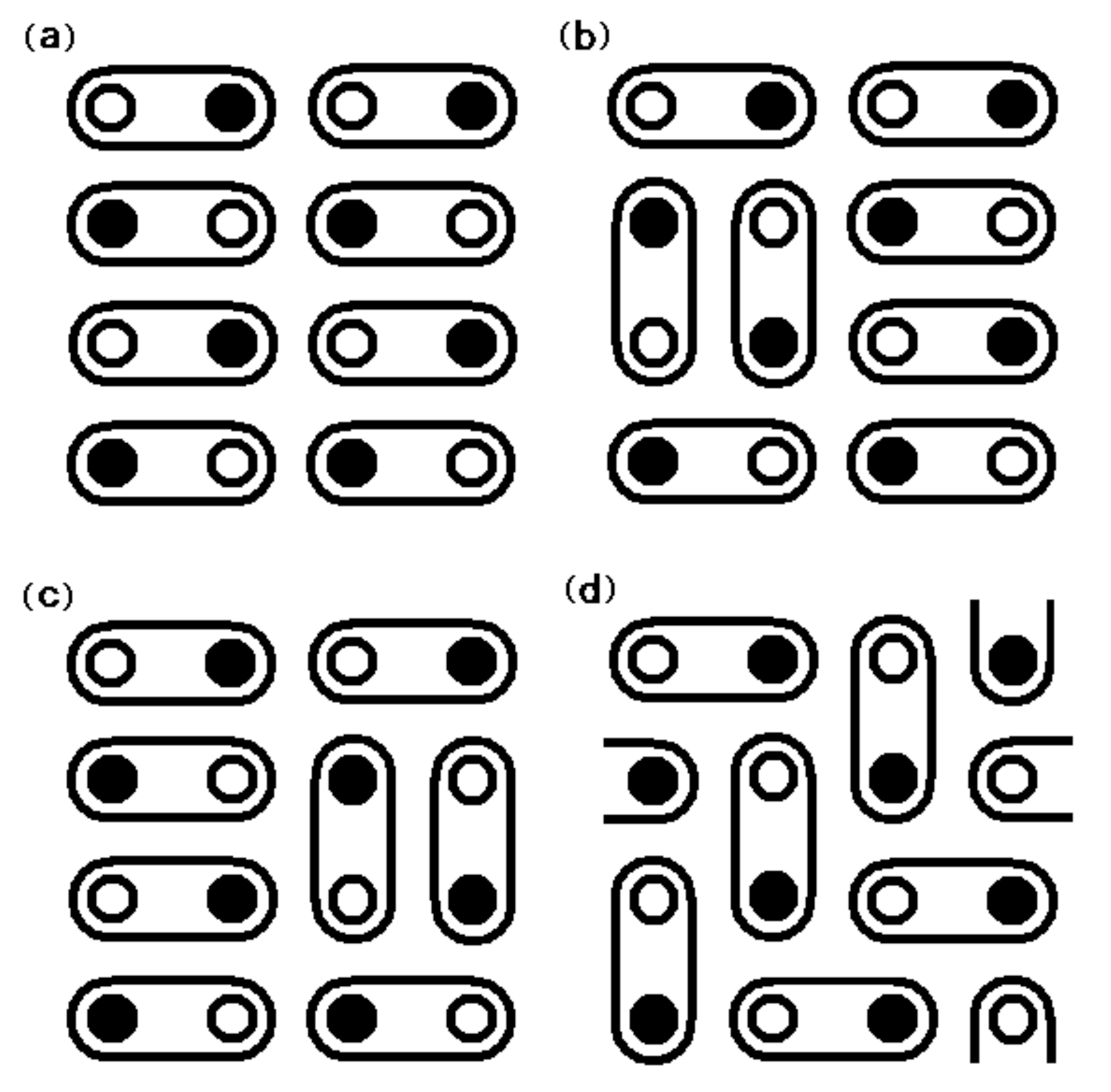}
\caption{Multitude of dimerized configurations for holographic glasses. (a)VBS-like configuration. (b)``Resonance excitation." (c)Another ``resonance excitation." (d)Generic configuration.}
 \label{7glass}
\end{figure}

Finally, let us consider the case of the square lattice, which is a special limit of the model considered in Sec.\ref{VBS} with $\para_0=1$. For illustration, let us take $N=10^{100}$ and $\sqrt{\lambda}=10$. Below the critical temperature $T_c\approx\frac{0.2}{a}$, the system again dimerizes through bubble nucleation. After such dimerization, ``spinon-pair excitation," corresponding to cutting a connected D5-brane into a pair of separated D5- and anti-D5-branes, is highly suppressed by a factor of $e^{-10^{100}}$ and thus the low temperature physics is dominated by dimerized configurations. But now there are infinitely many of them!  The simple VBS-like configuration [Fig.7(a)],  a ``resonance excitation" [Fig.7(b)] of that configuration, another ``resonance excitation" [Fig.7.(c)], and much more complicated configurations [like the one in Fig.7(d)] all have the same free energy at leading order in $N$, starting to differ only at order $N^0$. Furthermore, those configurations are connected only through highly suppressed tunneling events of brane recombinations in the bulk. Especially, note that even moving the ``resonance excitation" from site [Fig.7(b)] to site [Fig.7(c)] involves tunneling, and thus such excitations should not be thought of as usual quasiparticle excitations. This marks a stark difference from standard dimer models and more closely resembles glassy systems, with the low temperature physics dominated by a humongous number of metastable dimerized configurations. We also believe that this system does not have any long-range order: interactions between dimers are presumably mediated by light fields in the bulk, but, due to the peculiar geometric properties of AdS spacetime, such interactions are effectively rendered short ranged. Therefore, at large but finite $N$, we propose that the system transitions into a glassy phase at $T_c\approx\frac{0.2}{a}$, randomly
selecting one of the plethora of dimerized configurations as we cool it down.

\section{Future Directions}\label{conclusion}

In this paper, we engineered holographic lattices, holographic dimers, and holographic glasses.
In each instance, we chose to discuss a simple case to facilitate our analysis.
However, by varying the dials in these models, we can explore a large landscape of
possibilities, looking for qualitatively new phenomena.

First, we can freely dial various parameters that naturally appear in the system.
The number of D5-branes (anti-D5-branes) $N_f$ (${\bar N}_f$), their directions $v^{I}$, and the number of localized fermions $\nu N$ (${\bar \nu} N$) can be chosen independently for each site within a unit cell. Also, one can think of giving a mass to the localized fermions by
separating the D5-branes from the D3-branes in the radial direction, although the D5-brane configurations for such a case are analytically known only at zero temperature~\cite{CGS}. Turning on background magnetic fields~\cite{Magnetic} and/or chemical potentials for various $U(1)$s is another option, and we expect to see the dimerization transition at zero temperature. We should also mention that localized fermions can be replaced by localized bosons by replacing D5-branes by appropriate D3-branes~\cite{Ensemble,Boson}. The lattice structure can be chosen at will and does not even have to be periodic.

Second, we can consider more general background geometries of the form ${\rm AdS}_5\times X_5$ where $X_5$ is an arbitrary 5-dimensional Sasaki-Einstein manifold with volume of order $\lad^5$, following the spirit of~\cite{Sasaki}. For such studies, flux quantization conditions analogous to the one found in~\cite{Stable} need to be worked out for general $X_5$.
We can also explore other dimensions by, for example, considering ${\rm AdS}_4\times X_7$ and M2- or M5-branes in it, where $X_7$ is a 7-dimensional Sasaki-Einstein manifold.
Alternatively, we may also construct ``phenomenological models," abstracting out the essential ingredients of defects in AdS spacetime to focus on the applications of interest instead of
the details of the string theory engineering, following the spirit of recent investigations of Fermi liquids \cite{John}.

Third, we can try to engineer more elaborate systems with desired properties. In particular, it would be nice if we could make the fermions hop between the lattice sites so that we might have a holographic dual of the Mott transition. A conceptually straightforward but practically hard way to accomplish this is to cutoff our geometry at some radius or to consider $N$ D3-branes and a lattice of D5-branes on $M_4\times CY_6$ where $CY_6$ is a compact Calabi-Yau manifold, so that the D5-D5 strings are not completely decoupled. However it would be nice if we can add such hoppings without considering compact Calabi-Yau spaces~\cite{KKY}.
Once such hopping is added, if we can completely liberate localized fermions away from lattice sites in some phases but not others, we have a chance of having a holographic dual representation of transitions in which the volume of the Fermi surface jumps~\cite{Senthil}.

Turning to technical challenges, it would be interesting if we can develop reliable calculations
of the free energy at order $N^0$. At this order, the computation is plagued by bulk quantum effects, but the interaction between dimers should also become evident. This could lead to a deeper understanding of our mysterious holographic glasses by, among other things, giving
a small splitting between the different approximate ground states.

Finally, one of the most interesting handles on ground states of strongly correlated systems is
the entanglement entropy.  The elegant formulation of Ryu and Takayanagi allows its
computation in the gravity dual of a strongly coupled field theory, for simple enough choices of the boundary geometry partitioning the system into two subsystems \cite{Ryu}.  For bipartite lattice
systems, another useful notion of entanglement entropy, the ``valence bond entanglement entropy," which is a priori distinct from the von Neumann entropy,
was introduced in \cite{Alet}.  It may be interesting to consider such generalized notions of
entanglement entropy in our gravity duals of bipartite lattice systems.

\begin{acknowledgments}
We thank Stephen Shenker for very useful discussions.  S.K. is also grateful to Allan Adams, Oliver DeWolfe, and Igor Klebanov for important remarks.  This work was substantially initiated at the
KITP miniprogram  ``Quantum Criticality and the AdS/CFT Correspondence" in June of 2009, and we would like
to thank the organizers and participants for the very stimulating atmosphere they provided.   S.K. also acknowledges the hospitality
of the Aspen Center for Physics, and is supported by the NSF under Grants No. PHY-0244728 and No. PHY05-51164, and the DOE
under Contract No. DE-AC03-76SF00515.
A.K. is supported by the U.S.\ Department
    of Energy under Grant No.~DE-FG02-96ER40956.
S.Y. is supported by the Stanford Institute for Theoretical Physics and NSF Grant No. 0756174.
\end{acknowledgments}

\appendix

\section{Conventions}
\label{conventions}
We set $\hbar=c=k_B=1$.

Our conventions are basically those of~\cite{Polchinski2}, but we will exclusively work in the following frame, akin to Einstein frame:  $(g_{\mu\nu})_{\rm ours}\equiv g_s^{\frac{1}{2}}e^{-\frac{\Phi}{2}}(g_{\mu\nu})_{\rm string}$, $(B_{\mu\nu})_{\rm ours}\equiv (B_{\mu\nu})_{\rm string}$,  $(C_{(p)})_{\rm ours}\equiv g_s (C_{(p)})_{\rm string}$, and $(e^{\Phi})_{\rm ours}\equiv (e^{\Phi})_{\rm string}$. This frame will nicely segregate out factors of $g_s$. Explicitly, in our frame,
\bea
S_{\rm IIB}&=&\frac{1}{(2\pi)^7\alpha'^4 g_s^2}\int d^{9+1}x\sqrt{-g}[R-\frac{1}{2}(\partial_{\mu}\Phi)(\partial^{\mu}\Phi)-\frac{1}{2}\le(\frac{e^{\Phi}}{g_s}\ri)^{2}(\partial_{\mu}C_{(0)})(\partial^{\mu}C_{(0)})\nonumber\\
&&-\frac{1}{12}\le(\frac{e^{\Phi}}{g_s}\ri)^{-1}(H_{\mu\nu\lambda}H^{\mu\nu\lambda})-\frac{1}{12}\le(\frac{e^{\Phi}}{g_s}\ri)({\tilde F}_{(3) \mu\nu\lambda}{\tilde F}_{(3)}^{\mu\nu\lambda})-\frac{1}{480}({\tilde F}_{(5) \mu\nu\lambda\rho\sigma}{\tilde F}_{(5)}^{\mu\nu\lambda\rho\sigma})]\nonumber\\
&&-\frac{1}{2(2\pi)^7\alpha'^4 g_s^2}\int C_{(4)}\wedge H\wedge dC_{(2)}\\
{\rm with}&H&\equiv dB_{(2)},\ {\tilde F}_{(3)}\equiv dC_{(2)}-C_{(0)}H,\ {\tilde F}_{(5)}\equiv dC_{(4)}-\frac{1}{2}C_{(2)}\wedge H+\frac{1}{2}B_{(2)}\wedge dC_{(2)}
\eea
and $S_{\rm D5}=S_{\rm DBI}+S_{\rm CS}$ where the Dirac-Born-Infeld (DBI) action is given by
\be
S_{\rm DBI}=-\frac{1}{(2\pi)^5\alpha'^3g_s}\int_{\rm D5} d^{5+1}\xi\le(\frac{e^{\Phi}}{g_s}\ri)^{-1}\sqrt{-{\rm det}_{ab}\le[\le(\frac{e^{\Phi}}{g_s}\ri)^{\frac{1}{2}}g_{ab}+B_{(2)ab}+2\pi\alpha'{\cal F}_{ab}\ri]}
\ee
and the Chern-Simons (CS) action is given by
\be
S_{\rm CS}=\frac{1}{(2\pi)^5\alpha'^3g_s}\int_{\rm D5} \sum_{p=0,2,4,6}C_{(p)}\wedge e^{B_{(2)}+2\pi\alpha' {\cal F}},
\ee
with the understanding that $g_{ab}$, $B_{(2)}$, $C_{(p)}$, and $e^{\Phi}$ are all induced from background geometry.
Here, equations of motion that follow from varying the above action need to be supplemented by the self-duality condition on ${\tilde F}_{(5)}$, namely,
\be
*{\tilde F}_{(5)}={\tilde F}_{(5)}.
\ee

We will not explicitly write down appropriate Gibbons-Hawking surface terms.

\end{document}